\documentclass[twocolumn,prl,aps,superscriptaddress,showpacs]{revtex4}
\usepackage{dcolumn}
\usepackage{amsmath}
\usepackage{epsfig}
\begin{document}

% \thanks{Also at Physics Department, XYZ University.}%Lines break

\title{Resonant inelastic x-ray scattering study of holon-antiholon continuum in 
SrCuO$_2$}
\author{Young-June Kim}
\author{J. P. Hill}
\affiliation{Department of Physics, Brookhaven National Laboratory,
Upton, New York 11973}
\author{H. Benthien}
\affiliation{Department of Physics, Philipps University Marburg, D-35032,
Marburg, Germany}  
\author{F. H. L. Essler}
\affiliation{Department of Physics, Brookhaven National Laboratory,
Upton, New York 11973}
\author{E. Jeckelmann}
\affiliation{Institute of Physics, University of Mainz, D-55099 Mainz, Germany}
\author{H. S. Choi}
\author{T. W. Noh}
\affiliation{ReCOE and School of Physics, 
Seoul National University, Seoul 151-747, Korea}
\author{N. Motoyama}
\author{K. M. Kojima}
\author{S. Uchida}
\affiliation{Graduate School of Frontier Sciences, University of Tokyo, Bunkyo, Tokyo 113-8656, Japan}
\author{D. Casa}
\author{T. Gog}
\affiliation{CMC-CAT, Advanced Photon Source, Argonne National
Laboratory, Argonne, Illinois 60439}

\date{\today}

\begin{abstract}

We report a resonant inelastic x-ray scattering study of charge
excitations in the quasi-one-dimensional Mott insulator SrCuO$_2$. We
observe a continuum of low-energy excitations, the onset of which exhibits
a small dispersion of $\sim 0.4$ eV. Within this continuum, a highly
dispersive feature with a large sinusoidal dispersion ($\sim 1.1$ eV) is
observed. We have also measured the optical conductivity, and studied the
dynamic response of the extended Hubbard model with realistic parameters,
using a dynamical density-matrix renormalization group method. In contrast
to earlier work, we do not find a long-lived exciton, but rather these
results suggest that the excitation spectrum comprises a holon-antiholon
continuum together with a broad resonance.

\end{abstract}

\pacs{78.70.Ck, 78.30.-j, 71.10.Fd, 75.10.Pq}

\maketitle

%%%% introduction %%%%%%

The separation of spin and charge degrees of freedom is one of the most important
and fascinating properties of electrons in strongly correlated systems in one
dimension. In particular, it is well known that in the one-dimensional (1D)
Hubbard model, the low-energy physics is dominated by collective excitations of
decoupled charge and spin degrees of freedom called holons and spinons,
respectively \cite{Lieb68}. Experimentally, if one creates a hole by removing an
electron, this hole is expected to decay into a spinon and a holon, which can be
studied with angle resolved photoemission spectroscopy (ARPES) \cite{CKim}. The
situation is different for so-called ``particle-hole"  probes, such as optical
spectroscopy, resonant inelastic x-ray scattering (RIXS), and electron energy loss
spectroscopy (EELS). In these experiments, the total charge is conserved in the
scattering process, so that an electron is simply moved from one site to another,
creating a hole and a doubly occupied site. The decay of the hole creates a holon
and a spinon, while the double-occupancy decays into an antiholon and a spinon.  
Since photons and electrons strongly couple to the charge sector, the behavior of
holon-antiholon pairs can be studied with these particle-hole probes
\cite{Neudert98,Hasan02}.

The so-called corner-sharing chain cuprates Sr$_2$CuO$_3$ and SrCuO$_2$
are both charge-transfer insulators; that is, they have insulating gaps of $\sim
2$ eV arising from strong electron correlations. Since their crystal structure
is highly anisotropic, the electronic structure remains 1D over a wide
temperature range. Only at a very low temperature does magnetic order set in,
due to the small interchain coupling ($T_N \approx 2$ K for SrCuO$_2$
\cite{Matsuda97}). Based on the commonly-used measure of 
quasi-one-dimensionality, $T_N /J \sim 10^{-3}$, these compounds can be regarded as among
the best realizations of quasi-1D systems \cite{Steiner76}.  The quasi-1D nature
of both SrCuO$_2$ and Sr$_2$CuO$_3$ has been studied extensively with various
experimental techniques, including magnetic susceptibility \cite{Motoyama96},
ARPES \cite{CKim}, Raman scattering \cite{Misochko96}, optical
spectroscopy \cite{Popovic01}, EELS \cite{Neudert98}, RIXS \cite{Hasan02}, and
neutron scattering \cite{Zaliznyak03}.

In this Letter, we report a detailed study of the momentum dependence of
the low-energy charge excitations in SrCuO$_2$, utilizing the RIXS
technique. We observe a continuum of excitations arising from the creation
of particle and hole pairs.  Within this continuum, a well-defined
spectral feature with a large sinusoidal dispersion ($\sim 1.1$ eV) is
observed. We have also measured the optical conductivity
$\sigma(\omega)$ and carried out a dynamical density-matrix
renormalization group (DDMRG) calculation of the extended Hubbard model. A
comparison of the RIXS spectra with both the theoretical results and
$\sigma(\omega)$ yields a consistent picture of charge excitations in this
material: We find two dispersive spectral features, in contrast to the
earlier RIXS study \cite{Hasan02}. The first feature is the onset energy
of the holon-antiholon continuum, which exhibits small dispersion ($\sim
0.4$ eV). The second spectral feature is a more dispersive broad
resonance that is the remnant of the strong-coupling exciton. This result
contrasts earlier EELS work, in which the excitation spectrum was
interpreted as being dominated by a bound exciton at large momenta
\cite{Neudert98}.

In RIXS experiments, the incident x-ray energy is tuned to near the
absorption edge of the particular element of interest, so that the
inelastic scattering intensity of certain electronic excitations is
resonantly enhanced. In addition, this gives an element specificity that
is valuable in studying complex materials such as the cuprate compounds
\cite{Hasan02,Hill98,Abbamonte99,Hasan00,LCO-PRL}. The RIXS experiments
were carried out at the Advanced Photon Source on the undulator beamline
9IDB with the same setup used in previous studies \cite{LCO-PRL}.  The
scattering plane was vertical and the polarization of the incident x-ray
was kept perpendicular to the Cu-O plaquette. Note that the corner-sharing
CuO$_2$ chain runs along the {\bf c}-direction. A single crystal of
SrCuO$_2$ was grown using the traveling solvent floating zone method. The
crystal was cleaved along the $(0 \; 1 \; 0)$ plane and mounted on an
aluminum sample holder at room temperature in an evacuated chamber.

%==============================================================================
\begin{figure}
\begin{center}
\epsfig{file=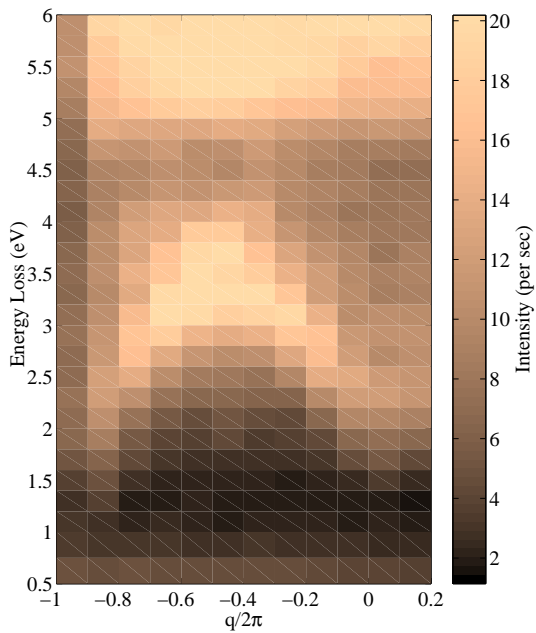,width=2.8in}
\end{center}
\caption{Pseudocolor plot of RIXS intensity as a function of momentum and 
energy transfer ($E_i=8982$ eV). } 
\label{fig1}
\end{figure}
%==============================================================================

In order to find the resonance condition, we first carried out a
study of the incident energy ($E_i$) dependence of the RIXS intensity.
There are two features that show resonance behavior in the energy-loss
($\omega \equiv E_i-E_f$) spectra, at $\omega \sim 3$ eV and $\omega \sim
6$ eV, respectively.  The intensity of the 3 eV feature shows large
enhancements at $E_i \approx 8982$ eV and $E_i \approx 8990$ eV, while the
6 eV feature is resonantly enhanced at $E_i \approx 8990$ eV and $E_i
\approx 8996$ eV. We note that these three resonance energies correspond
to the peaks in x-ray absorption spectra near the Cu K-edge
\cite{Tranquada91}. The intensity at 6 eV likely
arises from both the Cu K$\beta_5$ emission and a charge-transfer type
excitation commonly found in the RIXS studies of cuprates
\cite{Hill98,Abbamonte99}. Detailed studies of the $E_i$-dependence will be
reported elsewhere, and here we fix $E_i=8982$ eV, and focus on the
momentum dependence of the 3 eV feature.

Our main results are shown in Fig.~\ref{fig1}, in which the RIXS intensity
is plotted as a function of momentum and energy transfers. The pseudocolor
scale of the observed intensity is shown on the right side in units of
counts per second.  The momentum transfer is shown in reduced units of
$q/2\pi$ along the chain direction, and corresponds to the $(0\; 11\; q)$
position. Thus $q/2\pi=-0.5$ is the 1D Brillouin zone boundary, while
$q/2\pi=0$ is the zone center. The lower cut-off of $\omega =0.5$ eV shown
in Fig.~\ref{fig1}, is due to the large quasi-elastic contribution, and
$\omega=6$ eV is the upper limit of accessible energy transfer in this
setup. The plotted intensity is the raw RIXS intensity without any
absorption corrections.  Almost constant factors were obtained from three
different normalization methods \cite{Norm}, enabling us to compare the
absolute intensities between different momentum transfers directly, except
for $q/2\pi=-1$. The anomalously low overall intensity at $q/2\pi=-1$
arises from the very low angle of the incoming photons at this $q$.

%==============================================================================
\begin{figure}
\begin{center}
\epsfig{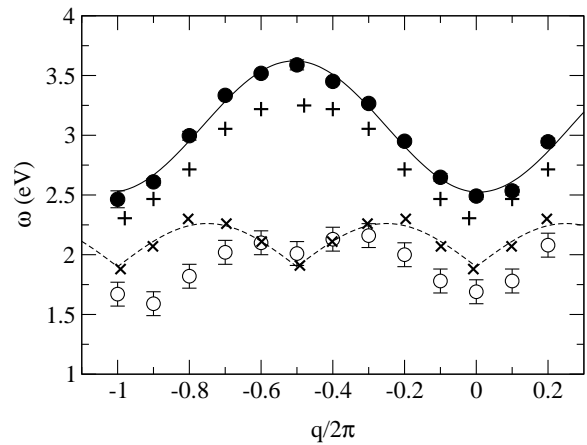}
\end{center}
\caption{
The dispersion relation of the peak position (filled symbol) and the onset
energy (open symbol) of the RIXS spectra. The solid line is
$\omega(q)=3.07-0.55 \cos(q)$. Calculated peaks and onset energies in
dynamical density-density correlation function, ${\cal
N}(q,\omega)$, are also plotted as plus and X symbols, respectively. The
dashed line denotes the spinon dispersion relation as described in the text.}
\label{fig2}
\end{figure}
%==============================================================================

\begin{figure}
\begin{center}
\epsfig{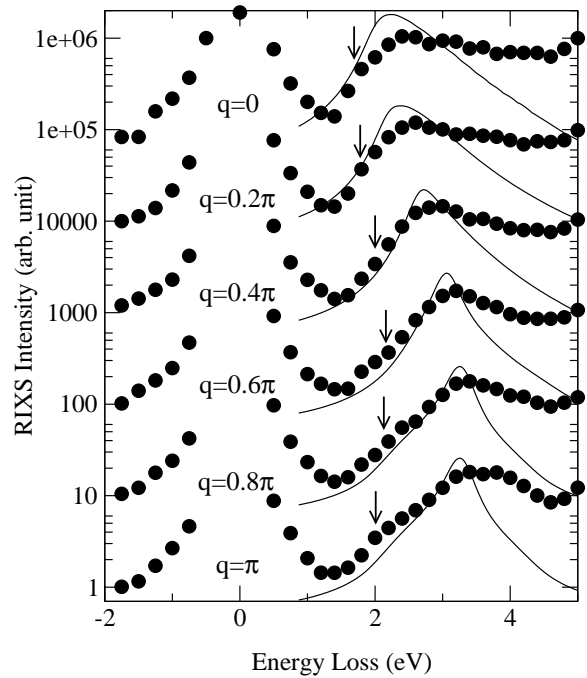}
\end{center}
\caption{Comparison of the RIXS spectra (circles) and ${\cal 
N}(q,\omega)$ (solid lines) at various momenta. To facilitate
a direct comparison, ${\cal
N}(q,\omega)$ is calculated with an experimental broadening of $\eta=0.5t$ 
and divided by $\sin^2(q/2)$. Arrows denote the onset of the RIXS 
spectral weight.}
\label{fig3}
\end{figure}

The most prominent feature of the RIXS spectra shown in Fig.~\ref{fig1} is
the highly dispersive feature at $\omega \approx 3$ eV. The momentum
dependence of the peak position of this feature, shown in Fig.~\ref{fig2},
exhibits a clear sinusoidal dispersion with a bandwidth of 1.1 eV.
However, closer inspection of Fig.~\ref{fig1} reveals the presence of
additional spectral weight on the low-energy side around $q/2\pi=-0.5$.
This low-energy intensity around the zone boundary is evident for the
$q=\pi$ data in Fig.~\ref{fig3}, where the scattered intensity is plotted
on a logarithmic scale. Since the momentum resolution at this position is
about 0.1 reciprocal lattice units, it is unlikely that
the additional low-energy intensity is due to a resolution effect.  Thus,
the best description of the observed spectra is that of a continuum of
excitations, in which the dispersive feature resides. Note that this
interpretation, particularly near $q=\pi$, is in contrast to previous EELS
and RIXS results. In their EELS study of Sr$_2$CuO$_3$, Neudert {\it et
al.} reported that the sharp spectral feature observed near $q=\pi$ is a
bound exciton mode, due to a strong inter-site Coulomb interaction
\cite{Neudert98}. The earlier RIXS study by Hasan {\it et al.} attributed
the dispersive feature to the Mott gap edge; that is, a lower bound of
particle-hole continuum \cite{Hasan02}. Our RIXS results suggest that the
continuum starts at much lower energy, and that the
particle-hole pairs do not form a bound exciton at $q=\pi$. The onset
energy of the spectral weight extracted from each scan is shown in
Fig.~\ref{fig3} as an arrow. The dispersion of these onset energies are
plotted in Fig.~\ref{fig2}. Note that the bandwidth of the onset energy
dispersion ($\sim 0.4$ eV) is much smaller than that of the peak position,
as discussed later.

\begin{figure} \begin{center}
\epsfig{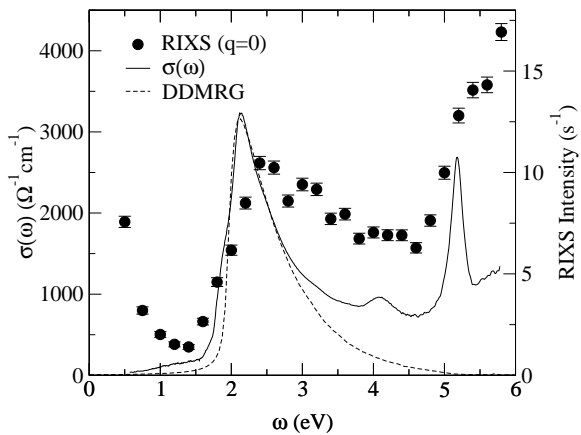}
\end{center} 
\caption{The optical conductivity $\sigma(\omega)$ at $T=90$ K is
plotted as a solid line, and compared with RIXS spectra obtained at the zone
center. The dashed line is the DDMRG 
calculation of $\sigma(\omega)$ with an experimental broadening of $\eta=0.1t$. 
} 
\label{fig4} 
\end{figure}

In concert with the RIXS data, we have also studied the optical properties of
SrCuO$_2$, measuring the reflectivity of a cleaved surface between 6 meV and 6
eV, with the polarization along the {\bf c}-axis. Optical conductivity,
$\sigma(\omega)$ calculated from the reflectivity data using the
Kramers-Kronig relation with an anchoring process, is plotted as a solid line in
Fig.~\ref{fig4} and compared with the $q=0$ RIXS data. The validity of the
Kramers-Kronig calculation was checked by directly measuring the complex
dielectric function with an ellipsometry method.  In the frequency range of
interest ($1 \sim 4$ eV), identical spectra were obtained from both techniques.
The $q=0$ RIXS data is expected to be different from that of 
$\sigma(\omega)=\omega {\rm Im}[\epsilon(0,\omega)]$, since
the response function measured in RIXS is believed to be close
to a dielectric loss function, Im$[-1/\epsilon(q,\omega)]$. Instead, one should
focus on the onset energy of the spectral weight, which should be independent
of both the particular response function being measured and matrix-element
effects. Both the RIXS data and the $\sigma(\omega)$ main feature exhibit onset 
energies around 1.7 eV, which corresponds to the continuum edge at 
the
zone center.

%==============================================================================
%==============================================================================

To understand the observed continuum and the dispersive spectral 
feature, we have studied a simple one-band half-filled 1D extended Hubbard 
model (EHM) \cite{Stephan96,Tsutsui00,Essler01,Jeckelmann03}, in which 
the nearest-neighbor
repulsion $V$ is explicitly included, in addition to the usual $t$ and 
$U$. It is clear that a full
theoretical description of ${\rm SrCuO_2}$ requires a multi-band model 
based on
the Cu $3d$ and O $2p$ orbitals. However, a precise and reliable determination
of dynamical correlation functions in such models is still a formidable
challenge. Several previous studies suggested that much of the essential physics
of both ${\rm Sr_2CuO_3}$ and ${\rm SrCuO_2}$ at low energies can be captured by
the EHM \cite{Stephan96,Neudert98,Tsutsui00,Essler01,Jeckelmann03}.  We have
therefore used a DDMRG algorithm \cite{Jeckelmann03} to determine the optical
conductivity and density response function in this model for realistic values of
$U$ and $V$ on chains of up to 128 sites. All previous studies either dealt with
unrealistic parameter values \cite{Neudert98} or suffered from severe
finite-size effects caused by the very small system sizes amenable to exact
diagonalization methods \cite{Hubsch01}.

The parameters $t$, $U$ and $V$ of the EHM were fixed by comparing DDMRG
calculations of the optical conductivity $\sigma(\omega)$ to the experimental
data and requiring compatibility with the value $J\approx 0.23$ eV for the
Heisenberg exchange coupling $J$ obtained from neutron scattering
\cite{Zaliznyak03}. As shown in Fig.~\ref{fig4}, an excellent fit for the
low-energy peak in $\sigma(\omega)$ is obtained for $t=0.435$ eV, $V/t=1.3$,
$U/t=7.8$. For these values we obtain $J\approx 0.24$ eV from both $1/U$
expansions and DDMRG computations of the spinon dispersion in the
single-particle spectral function. Note that the calculated $\sigma(\omega)$ in
Fig.~\ref{fig4} agrees with the measured spectra on an absolute scale.

%==============================================================================
%==============================================================================

The excitation spectrum of the EHM with these parameters consists of
scattering states of gapped, spinless charge $\mp e$ collective modes --
{\sl (anti)holons}, and gapless, chargeless spin $\pm 1/2$ collective modes
-- {\sl spinons}. In contrast to the strong-coupling results reported in
Ref.~\cite{Stephan96}, no excitonic excitations are found in any
part of the Brillouin zone. The current operator is charge neutral and hence
only excitations with equal numbers of holons and antiholons contribute to
$\sigma(\omega)$. Most of the spectral weight of the low-energy peak in
$\sigma(\omega)$ is attributed to excited states with one holon and one
antiholon each. The contributions of the spin degrees of freedom to
$\sigma(\omega)$ is small \cite{Essler01}.

The RIXS data is more difficult to model as at present there is no completely
satisfactory theory of RIXS. However, it is clear that the RIXS process involves
charge fluctuations that may be represented by the low energy excitations of the
EHM \cite{Abbamonte99,Hasan02,Tsutsui00}. We have therefore determined the
dynamical density-density correlation function ${\cal N}(q,\omega)$ by DDMRG.  
The results are shown in Fig.~\ref{fig3} as solid lines.  At small momenta
${\cal N}(q,\omega)$ is proportional to $q^2$ and thus we use a normalization
factor of $\sin^2(q/2)$ to compare ${\cal N}(q,\omega)$ at various momenta. We
note that \emph{on a lattice} the Fourier transform $V(q)$ of the Coulomb
potential is approximately given by $1/\sin^2(q/2)$. Therefore, the renormalized
${\cal N}(q,\omega)$ in Fig.~\ref{fig3} is probably close to the dielectric 
loss function
${\rm Im}[-1/\epsilon(q,\omega)] \propto V(q) {\cal N}(q,\omega)$ of the EHM. At
$q=0$ there is a broad ``band'' of states with a peak around 2.3 eV. As we move
towards the zone boundary, this feature narrows significantly.  Importantly,
however, even at $q=\pi$ the peak has an intrinsic width, that is, it is not a 
long-lived exciton mode. Furthermore, there is sizable
intensity below the peak at $q=\pi$. This is in contrast to the strong-coupling
$U\to\infty$ limit \cite{Stephan96}, where ${\cal N}(q\approx\pi,\omega)$ is
dominated by an excitonic holon-antiholon bound state. Our interpretation is
that as $U$ is reduced to smaller, more realistic values, the exciton acquires a
finite lifetime and turns into a broad holon-antiholon resonance.

Let us try to relate these findings to the RIXS data. In Fig.~\ref{fig3}, the
normalized ${\cal N}(q,\omega)$ is directly compared with the RIXS spectra at
equivalent $q$-positions.  The onset of the RIXS spectrum shown in Fig.~\ref{fig3}
seems to lie in the area where ${\cal N}(q,\omega)$ begins to show appreciable
spectral weight. The onset energy of ${\cal N}(q,\omega)$ is plotted in
Fig.~\ref{fig2} as a function of $q$ (crosses). The onset dispersion of ${\cal
N}(q,\omega)$ follows the dashed line, which is the single-spinon dispersion
relation shifted by a constant energy: $\frac{\pi}{2} J |\sin(q)| + 1.9$. This
spinon-like dispersion of the onset energy of ${\cal N}(q,\omega)$ is consistent
with the low-energy field theory prediction \cite{Controzzi02}. Qualitative
features of this spinon-like dispersion of the onset energy of ${\cal
N}(q,\omega)$ appears to be consistent with features of the RIXS data (open
circles, Fig.~\ref{fig2}), such as the bandwidth of $\sim \pi J/2$, and the
$q=\pi$ point being a local minimum. In addition, we have compared the dispersion
of the peak positions. As is indicated in Fig.~\ref{fig2}, the peak dispersion in
${\cal N}(q,\omega)$ is very similar to that of the RIXS data. However, such a
quantitative agreement between the RIXS peak positions and ${\cal N}(q,\omega)$
may well be coincidental, and further calculation of the full RIXS response
function is necessary to draw any firm conclusions.

In summary, we have studied the energy and momentum dependence of low-energy
charge excitations in quasi-1D SrCuO$_2$, using both optical spectroscopy and
resonant inelastic x-ray scattering.  We observe a continuum of excitations which
we associate with the creation of holon and antiholon pairs. The onset of the
holon-antiholon continuum exhibits spinon-like dispersion. Within this continuum,
a well-defined spectral feature with a large sinusoidal dispersion is also
observed.  By comparing the observed spectra with a DDMRG calculation of 1D EHM,
we interpret this well-defined feature as a broad holon-antiholon resonance due to
the inter-site Coulomb interaction. In contrast to earlier studies, we find that
the holon-antiholon pairs do not form an exciton, and the continuum starts at much
lower energy than previously reported.

%%%%% acknowledgement %%%%%

We would like to thank F. Gebhard and C. C. Homes for invaluable
discussions. The work at Brookhaven was supported by the US DOE, under
contract No. DE-AC02-98CH10886. Use of the Advanced Photon Source was
supported by the US DOE, under Contract No.  W-31-109-Eng-38. H.B.
acknowledges support by Optodynamics Center of the Philipps-Universit\"at
Marburg and thanks the Institute for Strongly Correlated and Complex
Systems at BNL for hospitality and support.


\begin{thebibliography}{10}

\bibitem{Lieb68}
E.~L. Lieb and F.~Y. Wu, Phys. Rev. Lett. {\bf 20},  1445  (1968).

\bibitem{CKim}
C. Kim {\it et~al.}, Phys. Rev. Lett. {\bf 77},  4054  (1996); 
Phys. Rev. B {\bf 56},  15589  (1997).

\bibitem{Neudert98}
R. Neudert {\it et~al.}, Phys. Rev. Lett. {\bf 81},  657  (1998).

\bibitem{Hasan02}
M.~Z. Hasan {\it et~al.}, Phys. Rev. Lett. {\bf 88},  177403  (2002).

\bibitem{Matsuda97}
M. Matsuda {\it et~al.}, Phys. Rev. B {\bf 55},  11953  (1997).

\bibitem{Steiner76}
M. Steiner, J. Villain, and C.~G. Windsor, Adv. Phys. {\bf 25},  87  (1976).

\bibitem{Motoyama96}
N. Motoyama, H. Eisaki, , and S. Uchida, Phys. Rev. Lett. {\bf 76},  3212
  (1996).

\bibitem{Misochko96}
O.~V. Misochko {\it et~al.}, Phys. Rev. B {\bf 53},  14733  (1996).

\bibitem{Popovic01}
Z.~V. Popovic {\it et~al.}, Phys. Rev. B {\bf 63},  165105  (2001).

\bibitem{Zaliznyak03}
I.~A. Zaliznyak, unpublished.

\bibitem{Hill98}
J.~P. Hill {\it et~al.}, Phys. Rev. Lett. {\bf 80},  4967  (1998).

\bibitem{Abbamonte99}
P. Abbamonte {\it et~al.}, Phys. Rev. Lett. {\bf 83},  860  (1999).

\bibitem{Hasan00}
M.~Z. Hasan {\it et~al.}, Science {\bf 288},  1811  (2000).

\bibitem{LCO-PRL}
Y.~J. Kim {\it et~al.}, Phys. Rev. Lett. {\bf 89},  177003  (2002).

\bibitem{Tranquada91}
J.~M. Tranquada {\it et~al.}, Phys. Rev. B {\bf 44},  5176  (1991).

\bibitem{Norm}
We considered normalization schemes based on elastic intensity, energy gain 
background ($\omega \ll 0$), and the fluorescence yield.

\bibitem{Stephan96}
W. Stephan and K. Penc, Phys. Rev. B {\bf 54},  R17269  (1996).

\bibitem{Tsutsui00}
K. Tsutsui, T. Tohyama, and S. Maekawa, Phys. Rev. B {\bf 61},  7180  (2000).

\bibitem{Essler01}
F.~H.~L. Essler, F. Gebhard, and E. Jeckelmann, Phys. Rev. B {\bf 64},  125119
  (2001).

\bibitem{Jeckelmann03}
E. Jeckelmann, Phys. Rev. B {\bf 67},  075106  (2003).

\bibitem{Hubsch01}
A. H\"ubsch {\it et~al.}, Phys. Rev. B {\bf 63},  205103  (2001).

\bibitem{Controzzi02}
D. Controzzi and F.~H.~L. Essler, Phys. Rev. B {\bf 66}, 165112 (2002).

\end{thebibliography}
\end{document}